\documentclass[showpacs,showkeys,floatfix,onecolumn]{revtex4}
\usepackage[cp1251]{inputenc}
\usepackage[english]{babel}
\usepackage[dvipdf]{color}
\usepackage{bm}
\usepackage{amsmath}

\begin{document}

\title{Comment on ``Thermodynamics of quantum crystalline membranes''}

\author{E.I.Kats and V.V.Lebedev}

\affiliation{Landau Institute for Theoretical Physics, RAS, \\
142432, Chernogolovka, Moscow region, Russia, and \\
Moscow Institute of Physics and Technology, \\
141700, Dolgoprudny, Moscow region, Russia}

\begin{abstract}

B. Amorim et al. \cite{AR14} reported the theoretical investigation of quantum crystalline membranes. In this comment we dismiss validity of their calculations based on a ``natural'' estimation of the ultra-violet (UV) divergent contributions into correlation functions. We claim that such calculations give qualitatively wrong results.

\end{abstract}


\maketitle

In the recent paper \cite{AR14} the authors study long-scale vibrational modes of free standing crystalline membranes. In particular, they examine the low temperature membrane areal thermal expansion coefficient and the membrane specific heat. The same systems (free standing crystalline membranes) were investigated in our paper \cite{KL14} where we demonstrated that quantum fluctuations produce logarithmic renormalization of the membrane elastic moduli regarding zero surface tension. However, the authors of the paper \cite{AR14} include into consideration a contribution into the surface tension (the term proportional to $k^2$ in the self-energy function) that comes from short-scale fluctuations. In our comment we warn that such terms cannot be calculated or even estimated via evaluation of the UV terms.

Let us remind the Landau-Wilson paradigm for applications of the field theory in condensed matter physics (see, e.g., the survey \cite{WK74} and the monographs \cite{ZJ89,SA99}). A macroscopic system can be described in terms of a few relevant variables (fields) possessing ``soft'' behavior, i.e., slow in time and smooth in space (in comparison with microscopic degrees of freedom). The macroscopic properties of the system can be analyzed in terms of a field theory that serves as a controllable theoretical tool for calculating correlation functions of the variables (fields). In theory, the correlation functions are presented as path integrals with a weight factor determined by the effective action or the Landau functional of the system. 

Usually, the correlation functions are calculated in the framework of a perturbation expansion. Then some ultraviolet (UV) divergent integrals appear unavoidably that cannot be calculated within the long-scale field theory. Such UV divergent terms have to be included into the renormalization (re-definition) of the factors of the Landau functional or of the effective action \cite{WK74,ZJ89}. This concept has frequently been misinterpreted, and it should not be mixed with renormalizability of the model Hamiltonian. The latter one is about a number of relevant interaction vertices.

What is worse that estimations of the UV divergent contributions, based on the ``natural'' microscopic cutoff (Debye frequency for the problem under consideration) are, generally, incorrect. The matter is that some macroscopic parameters have to be zero due to a symmetry or due to some other physical circumstances, whereas the ``natural'' estimations give nonzero values for the parameters. This is just the case for the freely suspended crystalline membranes that have zero surface tension $\sigma$. This fact follows from the equilibrium condition with respect to area variations for the freely suspended films that may adjust their area to minimize the film free energy.

One can say that $\sigma=\sigma_\mathrm{bare} + \sigma _{UV}=0$ where $\sigma_\mathrm{bare}$ is the ``bare'' value of the surface tension and $\sigma_{UV}$ is the UV (short-scale) contribution to the surface tension. Moreover, there is no way to calculate within the long-scale theory or to determine separately the contributions, $\sigma _{bare}$ and $\sigma _{UV}$. Only the sum of the both has physical meaning, and for tensionless free standing crystalline membranes $\sigma _{bare} + \sigma _{UV} =0$. It is obvious that in this situation the contribution $\sigma_{UV}$ cannot be used for estimating $\sigma$. Neither $\sigma _{UV}$ might be used to calculate thermodynamic characteristics. For example, flexular contribution into the low-temperature specific heat $C_p$ calculated at constant pressure (i.e., constant external stress) scales (up to logarithmic corrections) as $T^2$ and not linear in $T$ as in the paper \cite{AR14} with UV divergent term $\sigma _{UV}$ for the surface tension.

Next, the coefficient in front of UV divergent $k^2$ term in the self-energy function analyzed in the work \cite{AR14} cannot be interpreted as the renormalized bending modulus. It is the surface tension, as one can conclude by comparing the poles of the Green function and the phenomenological equation for bending fluctuations since the both should give the same long-wavelength dispersion law. The dispersion law reads as $\rho \omega ^2 = \sigma k^2 + \kappa k^4$, where $\rho $ is the mass density, $\kappa$ is the bending module and $\sigma$ is the surface tension. For tensionless free standing films $\sigma = 0$. If the membrane is somehow stretched then the surface tension $\sigma$ is non-zero. Then $\sigma$ is determined by an external force stretching the membrane, that is by a macroscopic factor. Then, evidently, $\sigma$ is much smaller than its ``natural'' estimation made in terms of the microscopic (atomic) parameters that determine $\sigma_{UV}$ .

It is worth to note certain similarity of the problem under discussion and long-scale description of continuous (second order) phase
transitions. The transition temperature $T_c$ is determined by microscopic material parameters and characteristics and cannot be found in the framework of the long-scale Landau-Wilson approach. Some UV divergent terms appearing at the renormalization procedure should be included into the definition of $T_c$: $T_c \to T_{c, bare} + T_{c, UV}$. However, neither $T_{c, bare}$ nor $T_{c, UV}$ can be found in the framework of the long-scale theory. Thus, in the long-scale theory $T_c$ has to be considered as a phenomenological parameter, its value can be taken from experimental data or microscopic calculations.

Note to the point (although it is a minor issue for our comment main message) that the authors of the paper \cite{AR14} state that in our work \cite{KL14} we neglect: {\it{(i) ``some relevant anharmonic terms''}} and {\it{ (ii) ``the effects of retardation''}}. In the statement (i) the authors of the paper \cite{AR14} have probably in mind the 4-th order interaction vertex for the bending fluctuations. We certainly do not neglect this vertex. The fact is that in the one-loop approximation the vertex appears only in the UV divergent (Hartree-like) contribution to the self-energy function, that has to be included into the redefinition of the surface tension (see above). In the statement (ii) the authors of the paper \cite{AR14} probably pointed to frequency independent renormalized interaction vertices in our work. The fact is that in our paper \cite{KL14} we exploit the standard renormalization group (RG) procedure that is valid if the dimensionless interaction constant is small. In this RG scheme the vertices remain frequency independent, indeed.

To conclude, we claim in this comment that the results obtained in the work \cite{AR14} on the basis of the ``natural'' UV estimation of the surface tension are qualitatively wrong, and are not applicable to free standing crystalline membranes.

We acknowledge support of Russian Science Foundation (grant 14-12-00475).

\end{document}